\documentclass{article}
\usepackage{graphicx} 
\usepackage{blindtext}
\usepackage{titlesec}
\usepackage{float}
\usepackage[usenames,dvipsnames]{color}
\usepackage[margin=3.5cm]{geometry}
\usepackage{bm}
\usepackage{amsmath}

\renewenvironment{abstract}
  {\par\textbf{\itshape Abstract}---\ignorespaces}
  {\par\medskip}
\newenvironment{keywords}
 {\par\textbf{\itshape Keywords}---\ignorespaces}
  {\par\medskip}

\title{Bayesian design and analysis of two-arm cluster randomised trials using assurance: extension to binary outcomes and comparison of MCMC and INLA}
\author{
  Abdullah Aloufi%
  \rlap{\textsuperscript{1}} \! \rlap{\textsuperscript{3}},
  Kevin J. Wilson%
  \rlap{\textsuperscript{1}},
  Nina Wilson%
  \rlap{\textsuperscript{2}},
  Lisa Shaw%
  \rlap{\textsuperscript{2}},
  Christopher Price%
   \rlap{\textsuperscript{2}}
}

\date{}

\begin{document}
\maketitle

\footnotetext[1]{School of Mathematics, Statistics \& Physics, Newcastle University, UK}
\footnotetext[2]{Population Health Sciences, Newcastle University, UK}
\footnotetext[3]{Department of Mathematics, Faculty of Science, Islamic University of Madinah, Saudi Arabia}

\begin{abstract}
\\ \textbf{Background/Aims:}
Bayesian designs for clinical trials using assurance to choose the sample size have been proposed in various trial contexts. Assurance allows for the incorporation of uncertainty on both the treatment effect and nuisance parameters into the sample size calculation. In the case of two-arm cluster randomised trials with continuous outcomes, assurance has been proposed with both a frequentist analysis (hybrid designs) and a Bayesian analysis (fully Bayesian designs). A Bayesian analysis in this context ensures a consistent treatment of probability throughout the design and analysis of the trial. In the fully Bayesian design inference has been achieved via Markov chain Monte Carlo (MCMC) sampling and, since assurance itself it evaluated via simulation, the result is a computationally intensive and often slow to run approach. In the case of two arm cluster randomised trials with binary outcomes assurance has not yet been explored to specify sample sizes, either in the hybrid or fully Bayesian case.

\textbf{Methods:}
This paper considers fully Bayesian designs for two-arm cluster randomised trials with continuous and binary outcomes. For the analysis of the trial we use a (generalised) linear mixed effects model. We summarise the inference for the treatment effect based on quantiles of the posterior distribution. We use assurance to choose the sample size. In the continuous case we investigate Integrated Nested Laplace Approximations (INLA) for inference to speed up calculation of the assurance, and compare INLA in computation time and accuracy to MCMC. In the binary case we develop the first fully Bayesian design for cluster randomised trials and conduct a similar comparison between INLA and MCMC. We demonstrate our novel approach using assurance to choose sample sizes for the SPEEDY cluster randomised trial, based on the results of a formal prior elicitation exercise with two clinical experts.\\
\textbf{Results:} We report comparisons of INLA and MCMC for a range of different scenarios for cluster RCTs, to determine when each inference scheme should be used, balancing the computational cost in terms of speed and accuracy. Overall MCMC with a very large number of samples produces very accurate inference, but does not scale well in terms of computational speed compared to INLA. Based on our simulation study, we recommend that INLA is used for inference in cluster trials with binary outcomes and large ($n>500$) cluster trials with continuous outcomes, and that MCMC is used in smaller ($n\leq 500$) cluster trials with continuous outcomes. Our case study demonstrated how to incorporate the uncertainty of trial clinicians into the sample size calculation to give an overall assessment of the likelihood of success of the trial. 
  
\textbf{Conclusions:}
A fully Bayesian design can be used for two arm cluster trials with both continuous and binary outcomes. INLA can allow for more efficient assessment of the assurance for cluster trials with binary outcomes and large cluster trials with continuous outcomes, without loss of accuracy in inference. A fully Bayesian design of a cluster randomised trial provides a coherent design and analysis framework and incorporates uncertainty in model parameters when choosing the sample size.
\end{abstract}


\sloppy

\section{Background/Aims}

\subsection{Introduction}
Sample size calculations are important in clinical trials as they balance the need for precision while taking into account practical considerations such as cost and time. It is unethical to recruit more participants than needed, but too few participants risks not being able to answer the research question, wasting time and money, and inconveniencing patients. In this paper, we focus on sample size calculations for two-arm superiority cluster randomised trials (CRT), both with a continuous outcome \cite{1} and with a binary outcome \cite{7}. 

For the CRT sample size calculation we will use a Bayesian approach, as it has the advantage of using prior knowledge, or information from previous studies, which is useful when there is uncertainty in the parameters and complexity in the inferential model. The Bayesian approach gives an intuitive interpretation in these cases. It also allows more flexible decision making. The Bayesian approach used to calculate the sample size is assurance \cite{1}, which is an alternative to power. The evaluation of the assurance typically requires a two-loop Monte Carlo scheme, sampling from a design prior distribution in the outer loop, and performing a Markov Chain Monte Carlo (MCMC) update to obtain samples of the treatment effect in the inner loop for each sample in the outer loop.

A particular challenge in this case, which is a problem more generally in Bayesian design of experiments, is computational cost. It can be time consuming to run a full MCMC scheme for every iteration in the Monte Carlo procedure described above. In an attempt to reduce computation time, in this paper we investigate Integrated Nested Laplace Approximations (INLA) \cite{9} as an alternative to MCMC \cite{10}. This approach has been considered for individually randomised controlled trials \cite{3}, but has not been investigated before in CRTs, which are more complex trials inferentially, requiring modelling of the cluster effects and intra-cluster correlation coefficient (ICC). \textcolor{black}{There are other papers that have discussed the comparison between INLA and MCMC \cite{i1,i2,i3,i4} with regression models of various types, but either their sole focus was accuracy, they only considered very large MCMC runs, or the models they considered were not comparable to those in this paper. Our investigation focuses on the trade-off between speed and accuracy of inference on the treatment effect based on approximation using INLA and MCMC under varying numbers of posterior samples. As such, it provides a new perspective on the relative merits of MCMC and INLA in a clinical trials context.}

We compare the inference resulting from MCMC using different numbers of posterior samples and INLA for continuous outcomes, considering a linear mixed effects model as in Wilson (2023) \cite{1}, defined in Section \ref{a}. In general, it should be faster to obtain the posterior distribution for the marginal treatment effect using INLA than using a sampling scheme such as MCMC, particularly for complex designs and large sample sizes. However, INLA is an approximation, whereas MCMC samples from the true posterior distribution, and so with enough samples can be arbitrarily accurate. We further outline Bayesian inference for a CRT with a binary outcome, and undertake a comparison of MCMC and INLA for this case. Based on our investigation, we provide guidance on when INLA and MCMC are most suitable for Bayesian analysis of CRTs.

We demonstrate the approach by calculating the sample size of the case study SPEEDY trial \cite{2} using assurance \cite{1}, for both continuous and binary co-primary outcomes. Based on our investigation we use MCMC for the continuous outcome and INLA for the binary outcome. To evaluate the assurance we use the prior distributions resulting from an expert elicitation exercise with the two co-leads in SPEEDY. We report the assurance and required sample sizes in each case from the priors for each expert, and from an equally weighted prior between the two experts. 

The paper is structured as follows. In Section \ref{Sec:power} we review a standard approach to power calculations for two-arm superiority CRTs for continuous and binary outcomes. In Section \ref{a} we detail Bayesian inference for two-arm superiority CRTs with continuous and binary outcomes. In Section \ref{assu} we detail how to calculate assurance for CRTs. In Section \ref{Sec:comp} we perform a simulation study comparing inference via MCMC and INLA in both cases, evaluating their accuracy and computation time. Section \ref{v} contains the application to the SPEEDY trial. We summarise the paper and identify future work in Section \ref{Sec:sum}.

\subsection{Power calculation for two-arm superiority CRTs}\label{c} \label{d}
\label{Sec:power}

Here we summarize standard power calculations for CRTs, to provide a contrast to the assurance described in Section \ref{a}.

The power for a two-arm CRT with a continuous outcome is given by the conditional probability that we reject the null hypothesis of a treatment effect of zero (for example), given an assumed treatment effect and values chosen for a set of nuisance parameters detailed below. We can approximate the power function, for sample size $n$ given by the product of the number of clusters $C$ and the average sample size in a cluster $\bar{n}$, for a one-sided Wald test of the treatment effect at significance level $\alpha$ \cite{4}, via,
\begin{equation}
P(n|\delta,\psi)= \Phi \left(\delta \sqrt{\frac{C (\bar n)}{4\sigma^2[1+\{(\nu^2+1)(\bar n)-1\}\rho]}}-z_{1-\alpha}\right)\label{power1}
\end{equation}
where $\delta$ is the treatment effect, $\psi=(\sigma,\rho, \nu)$ is the vector of nuisance parameters given by the overall standard deviation $\sigma$, ICC $\rho$ and coefficient of variation in cluster sizes $\nu$, $z_{1-\alpha}$ is the $100\times(1-\alpha)$ quantile of the standard Normal distribution, $\alpha$ is the significance level of the Wald test and $\Phi$ is the cumulative distribution function of the standard normal distribution. The sample size is chosen to be the smallest value which gives at least a desired \textcolor{black}{power $1-\beta$, where $\beta$ is the Type II error rate.}

The power in the binary case can be expressed \cite{6,13} as 
\begin{equation}
P(n\mid p_1,p_2,\rho) = \Phi \Biggl\{\frac{(p_2-p_1)-z_{1-\alpha}\sigma_p}{\sigma_D}\Biggl\}+ \Phi \Biggl\{\frac{(p_1-p_2)-z_{1-\alpha}\sigma_p}{\sigma_D}\Biggl\}
\label{power2}
\end{equation}
where $(p_1,p_2)$ are the probabilities of a positive primary outcome in the control and treatment arms respectively, $\sigma_p$ is the pooled standard deviation given by \[\sigma_p=\sqrt{\frac{\tau[\bar{p}(1-\bar{p})]}{n}}\]
and $\sigma_D$ is the standard deviation of the difference between the probabilities, and is given by
\[\sigma_D=\sqrt{\frac{2\tau[p_1(1-p_1)+p_2(1-p_2)]}{n}}.\]
Here $\bar{p}=(p_1+p_2)/2$ and $\tau=1+ \rho(\bar{n}-1)$ is the design effect, which is assumed equal in the control and treatment arms. We choose $n$ as the smallest value that gives the required \textcolor{black}{power $1-\beta$}. \textcolor{black}{In both formulas for the continuous and binary outcomes if you use $\frac{\alpha}{2}$ in place of $\alpha$ that will give the power for the two sided test.}

In general, we will have uncertainty about the true values of the (nuisance) parameters in the power calculations above. By defining a prior distribution on the (nuisance) parameters, rather than assuming single values as in power, we can take this uncertainty into account in the sample size calculation. The resulting quantity is known as the assurance, and can be used to choose the sample size for a CRT in combination with either a frequentest or a Bayesian analysis, respectively known as a hybrid and a fully Bayesian design. 

\section{Methods}

\subsection{Bayesian inference for two-arm CRTs} \label{a}

An alternative to the hypothesis-testing analyses which formed the basis of the power functions in the previous section is to perform a Bayesian analysis of the trial. This has the advantage of allowing prior information to be incorporated into the analysis, and provides a coherent framework for design and analysis if the assurance is to be used to choose the sample size, which will be described in Section \ref{assu}. In this section, we detail Bayesian inference for CRTs. 

We describe the inference for the treatment effect for a CRT with a continuous outcome, based on the posterior distribution, as described in Spiegelhalter (2001) \cite{5}. For the binary outcome we can perform inference using a similar approach to that of Turner (2001) \cite{7}. Then, based on this inference, we use the developed assurance from Wilson (2023) \cite{1} in Section \ref{assu} to choose the CRT sample size. For the inference we  consider comparison of treatment with control. 

A (generalised) linear mixed-effects model can be used, with continuous response $Y_{ij} \sim N(\mu_{ij},\sigma^2_w)$ or binary response  $Y_{ij} \sim Bern(\theta_{ij})$, where $Y_{ij}$ are observed for individuals $i=1,...,n_j$ in clusters $j=1,...,J$, and the linear predictor is given by 
\[\eta_{ij}=\lambda+X_j\delta+c_j, \quad c_j \sim N(0,\sigma^2_b)\]
with $\eta_{ij}= \mu_{ij}$ for the continuous outcome and $\eta_{ij}= \log(\frac{\theta_{ij}}{1-\theta_{ij}})$ for the binary outcome. In addition, $\lambda$ is the control arm mean response, $X_j=1$ if cluster $j$ is the treatment arm and $X_j=0$ otherwise, $\delta$ is the treatment effect and $c_j \sim N(0,\sigma^2_b)$ is a random cluster effect, with $\sigma^2_b$ being the between cluster variance, with additionally $\sigma^2_w$, the within-cluster variance in the continuous case. 

For Bayesian inference the parameters $\bm\Psi =(\lambda,\delta,\sigma^2_b)'$ and possibly $\sigma ^2_w$ require prior distributions. There are various possibilities, but suitable forms for the marginal prior distributions \cite{1,5} are
\[\lambda\sim N(m_\lambda, v_\lambda), \quad \delta\sim N(m_\delta, v_\delta),\] 
\[\tau_b=\frac{1}{\sigma^2_b}\sim \Gamma(r_b,s_b),\quad \tau_w=\frac{1}{\sigma^2_w}\sim \Gamma(r_w,s_w),\]
where each $(m,v)$ and $(r,s)$ are hyper-parameters to be chosen. In the analysis at the end of the trial, we may choose to make these prior distributions relatively non-informative, to be consistent with equipoise. 

The inference in both cases are not conjugate, and so numerical or approximation methods are needed to evaluate the posterior distribution on the treatment effect $\delta$. 



Previous work \cite{1,5} in the continuous case has considered simulation from the posterior distribution of the treatment effect using MCMC. For large or complex CRTs this can be computationally costly, and, when many runs of the MCMC are required as described for the design of the trial in Section \ref{assu}, it may not be feasible to use MCMC at all. We propose INLA as an alternative to MCMC for inference on the treatment effect in CRTs, and will compare MCMC and INLA under various scenarios, focusing on their accuracy and computational cost.

In the analyses in this paper we perform inference via MCMC using the R package \verb|rjags|  \cite{10}. The rjags package is used for Bayesian data analysis and interfaces between R and the JAGS library \cite{12}. It uses a combination of Gibbs sampling, Metropolis-Hastings sampling and slice sampling to sample from the posterior distribution. In our implementation of MCMC in \verb|rjags| we use a burn-in period to allow the MCMC chains to converge before recording samples. 

To perform inference using INLA, we use the \verb|INLA| package from the R-INLA project \cite{9}. The idea behind INLA is that it approximates the required integral to evaluate the posterior distribution using Laplace's method. It can be used for the analysis of CRTs since the (generalised) linear mixed effects models can be written as latent Gaussian models, for which the Laplace method can be applied. For further information see Gómez-Rubio (2020) \cite{INLA}. We obtain the required quantiles from the posterior distribution of the treatment effect directly from INLA, without the need for sampling. 


\subsection{Assurance}\label{assu}
Following Wilson (2023) \cite{1}, assurance evaluates the unconditional probability that the trial finds a significant treatment effect. This allows an appropriate sample size choice in the planning of any cluster RCT, and is not conditional on chosen values of unknown parameters in the same way as the power. 

Define an event ``Success" to be the successful outcome of the CRT, i.e., treatment is superior to control. Then, for the sample size $n$, the assurance is given by    
\begin{align}
A(n) 
       & = \iint I_A[\textrm{Success}|\bm y]f(\bm y|\Psi,n)\pi_D(\Psi)d\Psi d\bm y \nonumber,
\end{align}
where $\bm y$ is the vector of responses, $\Psi$ is the vector of model parameters, $I_A[\textrm{Success}|\bm y]$ is an indicator function which takes the value 1 if the trial results in a success, $f(\bm y|\Psi,n)$ is the probability density function of $\bm y$ and $\pi_D(\Psi)$ is the design prior distribution for $\Psi$.

The total sample size in a cluster RCT is given by $n= \sum^J_{j=1}n_j$. Specifying a total sample size in place of each individual cluster sample size is standard practice in cluster RCTs. In the case where there will not be the same number of individuals in each cluster, we can model the number of individuals in each cluster $\bm n=(n_1,\cdots,n_J)'$ as \[\bm n \sim Multinomial (\textcolor{black}{n},\bm p)\]
where $\bm p=(p_1,...,p_J)'$ and $p_j$ is the random selection probability of an individual coming from cluster $j$. Similar to Wilson (2023) \cite{1} we choose for $\bm p$ a symmetrical Dirichlet prior distribution, $\bm p \sim Dirichlet(\bm a)$, in the case where we have no reason to think any particular cluster is likely to be larger than any other {\em a priori}. In this case $\bm a=(a_1,...,a_J)'$, and $a=a_1=\cdots=a_J$. When the values of $a$ are smaller the variation in cluster sizes will increase. When $a_j\neq a_{j'}$ for $j \neq j'$ this will lead to unequal prior probabilities of recruitment in each cluster.  

The assurance for total sample size \textcolor{black}{$n$} can be evaluated using a standard Monte Carlo simulation approach, as
\[A(n)=\frac{1}{L} \sum^L_{\ell=1}I(\textrm{Success}|\bm y^{(\ell)}),\]
where ``Success'' denotes that treatment is found superior to control based on the posterior distribution in the analysis of the CRT and $I$ is an indicator variable which takes the value 1 if this is true. To obtain this we sample $(\Psi,\bm p)^{(\ell)}=(\alpha,\delta,\sigma_w,\sigma_b,\bm p)^{(\ell)}$ in the continuous case or $(\Psi,\bm p)^{(\ell)}=(\alpha,\delta,\sigma_b,\bm p)^{(\ell)}$ in the binary case from the design prior distribution, for $\ell= 1,...,L$, and then, based on these values, we sample $\bm n^{\ell}$ from the multinomial distribution and $\bm y^{(\ell)}$ from the likelihood function. Based on this synthetic trial data we evaluate the posterior distribution based on the analysis prior distribution. 

In the case of MCMC we obtain samples of $\delta^{(\ell)}\mid\bm y^{(\ell)}$ and assess if a required quantile is above zero (or perhaps above the MCID) empirically to evaluate the indicator function. This results in a two-loop sampling scheme. We denote the samples of $\delta$ in this inner loop using subscript $k$, i.e., $\delta^{(k\ell)}$ for $k=1,\ldots,K$. In the case of INLA, we can obtain the approximation of the required quantile of $\delta$ directly, with no additional inner loop sampling. For a chosen sample size, these approximations will provide the assurance based on a total number of samples of $L \times K$ or $L$ respectively, excluding the burn-in iterations in the MCMC and the approximation calculations in INLA.

\section{Results}
\subsection{Comparison of INLA VS MCMC}
\label{Sec:comp}

For both the continuous and binary outcome we simulate a CRT \textcolor{black}{with two different numbers of clusters; $C=8$ and $C=12$,} and choose the following true values for the parameters,
$\alpha=1$ and $\delta=2$. The value of the intercept is arbitrary, and different intercept values do not affect the reported results. The precisions are \textcolor{black} {$\tau_b=\frac{1}{\sigma^2_b}=\{5,10\}$, and  $\tau_w=\frac{1}{\sigma^2_w}=\{0.25,0.01\}$ for the continuous outcome, which gives two different ICC values of $\rho =\{0.05,0.01\}$, representing moderate and relatively strong intra-cluster correlations in a CRT}. \textcolor{black}{Therefore, we have 4 different simulation scenarios for the comparison; considering 8 and 12 clusters with ICC values of 0.01 and 0.05.} 

To compare MCMC to INLA we consider a range of numbers of MCMC samples, $K$, from ``small'' runs to ``large'' runs, specifically $K=\{100, 1000, 10000\}$. \textcolor{black}{ We also used $K=100,000$ for one scenario ($C=8$ clusters with $\rho=0.05$), and decided not to include it for the other scenarios as it was very slow and gave almost identical results to when $K=10,000$}. We vary the sample size in the simulated hypothetical trial $N=\{100,500,1000,2000,10000\}$, and record the time in seconds to obtain the posterior distribution and the accuracy of the inference, evaluated as the difference between the posterior median of the treatment effect and its true value. We repeat the simulation of each hypothetical trial 100 times and report the mean values and standard deviations of these two quantities.

\begin{figure}[!h]
    \centering
\graphicspath{ {./images/} }
\includegraphics[width=\linewidth]{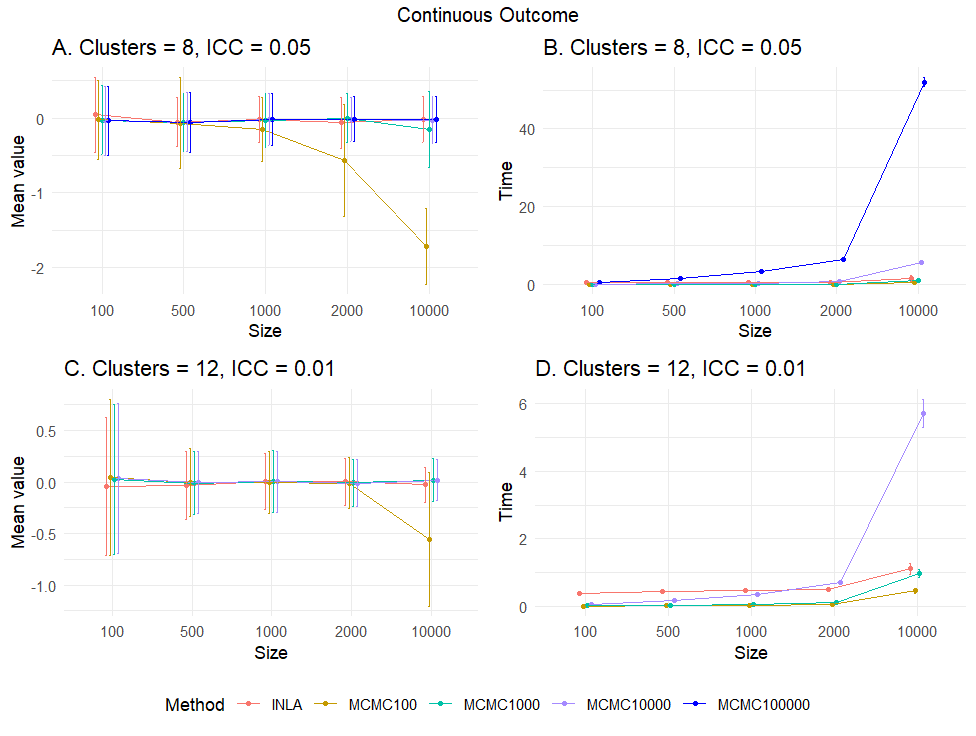}
\caption{The difference between the posterior median and the true treatment effect and the run time for each method, in the continuous \textcolor{black}{outcome case for scenarios $(C=8,\rho=0.05)$ and $(C=12,\rho=0.01)$ under each total  sample size}. In (A) MCMC with $K>100$ and INLA both are accurate in the \textcolor{black}{scenario with $C=8$ clusters and an ICC of $\rho=0.05$}. However, when using $K=100$ MCMC samples considering sample sizes of $N=2,000$ and $N=10,000$ the result is not accurate, as the posterior seems not to converge due to the small number of MCMC samples. INLA appears be as accurate as MCMC with $K=10,000$ MCMC samples. \textcolor{black}{In (C), for the scenario $C=12$ and $\rho=0.01$, MCMC is more accurate overall than the MCMC in (A). Also, when $K=100$ with sample sizes of $N=2000$ and $N=10000$ the inference is much improved.} In (B) \textcolor{black}{and (D)} MCMC run time is generally faster than INLA when the sample size is small. However, INLA scales better to large sample sizes. \textcolor{black}{In addition, the computation time does not differ much when using different ICC values, but increases substantially when increasing the number of clusters.}  }
\label{fig:mesh1}
\end{figure}

\begin{figure}[!h]
    \centering
\graphicspath{ {./images/} }
\includegraphics[width=\linewidth]{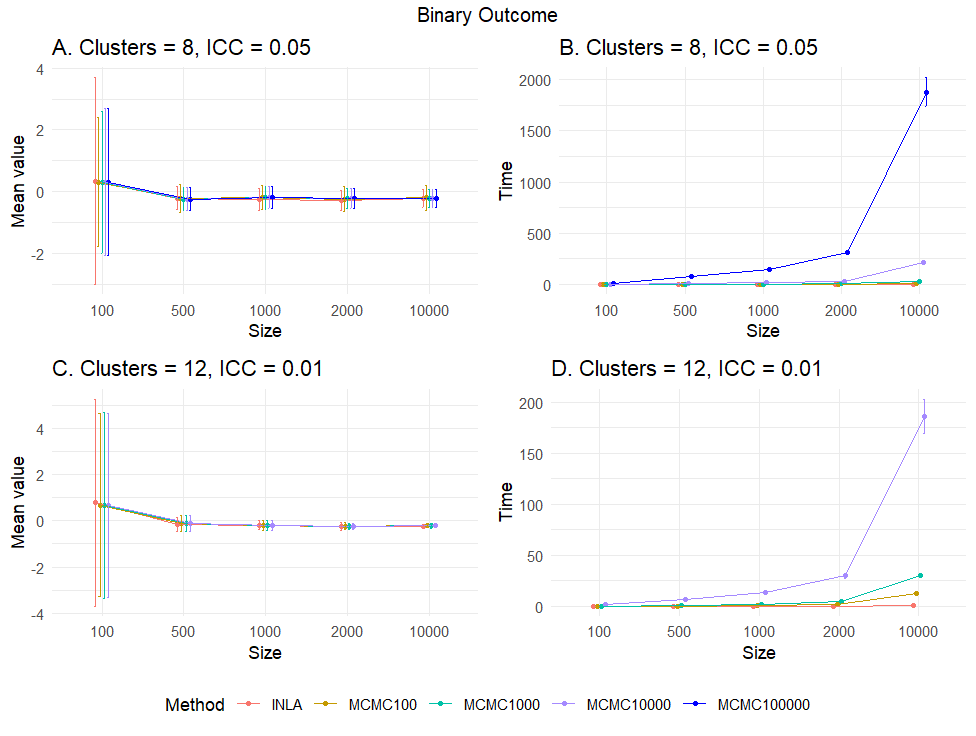}
\caption{The difference between the posterior median and the true treatment effect and the run time for each method in the binary \textcolor{black}{outcome case for scenarios $(C=8,\rho=0.05)$ and $(C=12,\rho=0.01)$ under each total  sample size}. In (A), \textcolor{black}{for 8 clusters with an ICC of 0.05} and with a sample size of $N=100$ the result is not consistently accurate for any of the inference methods. With binary data there is not enough information for accurate inference with such a small sample size. However, for larger sample sizes the estimation of the treatment effect is accurate for INLA and all of the different numbers of MCMC samples. The accuracy of INLA is consistently between the accuracy of MCMC with $K=10,000$ and $K=100,000$ MCMC samples. \textcolor{black}{Similarly, in (C) for 12 clusters with an ICC of 0.01, the result is accurate except when using $N=100$, and the uncertainty decreases when using $N\geq1000$, as a result of the low ICC value of 0.01. } For \textcolor{black}{(B) and} (D) MCMC with a large number of MCMC samples runs very slowly. Therefore, using INLA in general for the binary outcome case is useful. }
\label{fig:mesh2}
\end{figure}

See \textcolor{black}{ Figure \ref{fig:mesh1}(A, C) and Figure \ref{fig:mesh2}(A, C)} for the reported mean values of the posterior median minus $\delta$ and \textcolor{black} {Figure \ref{fig:mesh1}(B, D) and Figure \ref{fig:mesh2}(B, D)} for the time to obtain the posterior distribution, for the continuous and binary outcomes respectively. In each case we include both the mean and an approximate 95\% interval, the mean plus and minus two standard deviations.

Overall, in the continuous outcome case, we see that both INLA and MCMC are accurate for small trial sample sizes, with MCMC requiring at least 10,000 samples from the posterior distribution for trials with large sample sizes to ensure convergence. MCMC is faster than INLA for small sample sizes, but INLA is much faster than MCMC for large CRTs. This suggests that we should use MCMC with at least 10,000 samples to analyze continuous outcome CRTs with sample sizes of 100-500, and INLA for CRTs with a sample size above 500. \textcolor{black}{In addition, as we increase the number of clusters to $C=12$ and reduce the ICC to $\rho=0.01$, the result tends to be more accurate, even when using a small number of MCMC samples - this makes intuitive sense as both of these changes increase the effective sample size. The results for the remaining two scenarios, $\{(C=8,\rho=0.01),(C=12,\rho=0.05)\}$ are given in the supplementary materials. } 

For the binary case INLA is as accurate as MCMC with a large number of posterior samples for all CRT sample sizes, and is considerably faster. In the binary case, MCMC is not able to exploit the same conjugacy in the precision priors as the continuous case, explaining this disparity. The result is that INLA is a suitable approach to use for inference for two-arm cluster RCTs with a binary outcome, irrespective of the sample size of the CRT.

\textcolor{black}{Based on the simulations in each of the four scenarios we provide the following conclusions for the fastest approaches that provide accurate Bayesian inference in two-arm cluster randomised trials:
\begin{itemize}
    \item For a continuous outcome we found MCMC with $10,000$ posterior samples to be best when the total sample size is small (generally less than 1000) and INLA to be best when the total sample size is large.
    \item For a binary outcome we found INLA to be best for all total sample sizes. 
\end{itemize}}

\subsection{Application to the SPEEDY trial}\label{v}

\subsubsection{Introduction to SPEEDY}

 SPEEDY \cite{2} is a two-arm CRT which aims to determine the clinical and cost effectiveness of a novel specialist prehospital redirection pathway intended to facilitate thrombectomy treatment for acute stroke compared to standard care. The study has co-primary outcomes of thrombectomy rate and time to thrombectomy. The unit of randomization is the ambulance station and the sample size for the time to thrombectomy outcome is 564 participants and for the thrombectomy rate outcome is 894 participants. The primary analysis population is ambulance suspected stroke \textcolor{black}{who met the pathway initiation criteria and who were diagnosed with ischemic stroke following hospital assessment}, which is a subset of the full study population.

 The sample size for time to thrombectomy is based on 90\% power, $\alpha=0.05$, the one sided significance level, $\delta=30$ minutes as a reasonable smallest clinical meaningful difference for the time to thrombectomy between the arms, 150 clusters allocated 1:1 to the two arms, $\rho=0.01$ based on \cite{k1,k2}, the ICC and $\sigma=120$, the standard deviation of the time to thrombectomy in minutes. In terms of the power calculation detailed in Section \ref{c}, the value used for the coefficient of variability in cluster size was $\nu=0$, as cluster size variability was not considered in the sample size calculation. The required average cluster size can then be found from (\ref{power1}), and then multiplied by the total number of clusters to give the required sample size.
 
 Similarly, for the sample size calculation for the thrombectomy rate, the same values of power, significance level $\alpha$, the number of clusters, cluster allocation and ICC were used with, additionally, assumed rates of $p_1=0.132$ and $p_2=0.216$. Based on (\ref{power2}), we find the required sample size. 

We will use the SPEEDY trial \cite{2} as a case study to demonstrate the Bayesian CRT design. 

\subsubsection{Elicitation for the SPEEDY Trial}
In line with standard frequentest sample size calculations, the SPEEDY trial did not account for uncertainty in the model parameters. We wish to incorporate such uncertainty by using the assurance in place of power. This requires informative design prior distributions for each model parameter. We used expert elicitation to determine suitable prior distributions for the SPEEDY trial parameters, relating elicited values on observable quantities to the design prior distributions of interest. We will use these design prior distributions in our assurance calculation in the next section. 

To perform the elicitation, we first prepared an evidence dossier for the quantities of interest. We held an elicitation workshop with two experts who are the co-leads of the SPEEDY trial. In this elicitation workshop we used the quartile method to perform individual elicitations of the quantities of interest. However, we did not elicit the cluster size variability $\nu$ in the session as the experts felt that this would be better specified based on existing data. Instead we specified this prior based on the number of staff at each of the ambulance stations in SPEEDY, assuming that this would be proportional to the number of patients they would recruit in the trial. The elicitation approach we used was a variation on the Sheffield Elicitation Framework, detailed in \cite{8,11}. Full details of the elicitation and the documentation used are provided in the Supplementary Material.

\subsubsection{Assurance for the time to thrombectomy}\label{b}

\begin{figure}[H]
    \centering
\graphicspath{ {./images/} }
\includegraphics[width=\linewidth]{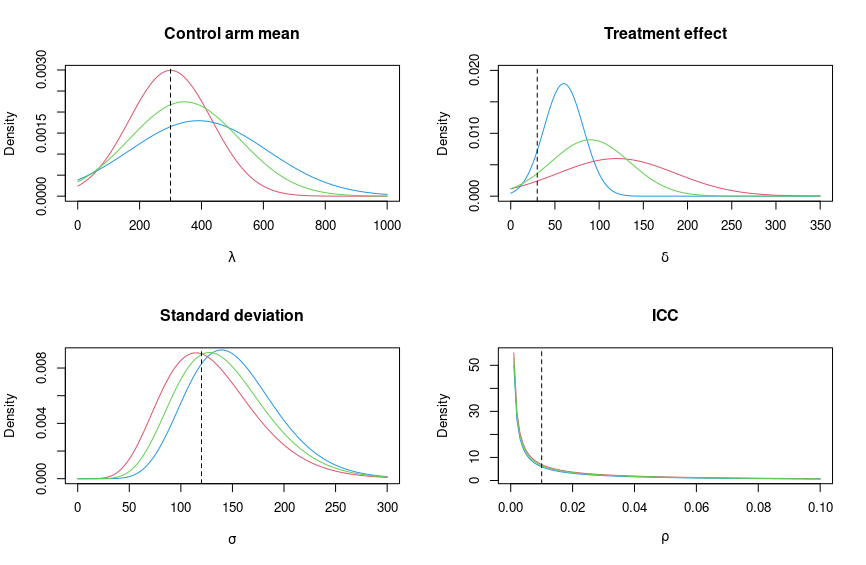}
\caption{The elicited prior probability density functions for the parameters  ($\lambda, \delta, \sigma, \rho $), for time to thrombectomy. Expert 1 is given in red, expert 2 in blue and the average in green. The vertical dashed lines are the values used in the original power calculation.}
 \label{fig:den}
\end{figure}

We reproduce the sample size calculation for time to thrombectomy using the assurance, as detailed in Section \ref{assu}. \textcolor{black}{Based on the general advice from the results of the simulation study in Section \ref{Sec:comp} we use MCMC for inference with $K=10,000$ samples}. To do so, we need to define the design prior distribution on the model parameters based on the elicitation results. We have three different sets of elicited design prior distributions using the information from expert 1, expert 2 and an equally weighted average of both experts' distributions. The priors resulting from the elicitation for experts 1 and 2, and the average, for $\lambda$, $\delta$, $\nu$, $\sigma$ and $\rho$, for time to thrombectomy, are given in Table 1. The marginal prior distributions for $\lambda$, $\delta$, $\rho$ and $\sigma$ are also provided in Figure \ref{fig:den}.

\begin{center}
\begin{table}[H]
\centering
\caption{The elicited prior distributions for each expert, and the average of both, for time to thrombectomy \textcolor{black}{and thrombectomy rate}. $\Gamma$ represents the gamma distribution and $B$ represents the beta distribution. \textcolor{black}{The prior distributions for $\nu$ and $\rho$ are used in both cases.} }
\begin{tabular}{ |c|c|c|c| } 
\hline
& Expert 1 & Expert 2 & Average \\
\hline
$\nu$ &$\Gamma(0.48,0.16)$ & $\Gamma(0.48,0.16)$ &$\Gamma(0.48,0.16)$\\
\hline
$\rho$& $B(0.10,2.1)$ & $B(0.08,2.1)$ & $B(0.09,2.1)$\\
\hline
\multicolumn{4}{|c|}{\textcolor{black}{Time to thrombectomy}} \\
\hline
$\lambda$& $N(300,133.4^2)$ & $N(390,222.4^2)$ & $N(345,177.9^2)$\\ 
\hline
$\delta$ &$N(120,66.7^2)$ & $N(60,22.3^2)$& $N(90,44.5^2)$\\ 
\hline
 $\sigma$&$\Gamma(7.99,0.06)$ & $\Gamma(11.73,0.08)$& $\Gamma(9.68,0.07)$\\
\hline
\multicolumn{4}{|c|}{\textcolor{black}{Thrombectomy rate}} \\
\hline
$\lambda$& $N(-1.22,0.45^2)$ & $N(-1.99,0.35^2)$ & $N(-1.64,0.42^2)$\\ 
\hline
$\delta$ &$N(0.54,0.36^2)$ & $N(0.6,0.19^2)$& $N(0.58,0.28^2)$\\ 
\hline
 $\sigma_b$&$\Gamma(0.14,0.63)$ & $\Gamma(0.11,0.39)$& $\Gamma(0.12,0.47)$\\
\hline
\end{tabular}
    \label{tab:1}
\end{table}
\end{center}

\begin{figure}[H]
    \centering
\graphicspath{ {./images/} }
\includegraphics[width=\linewidth]{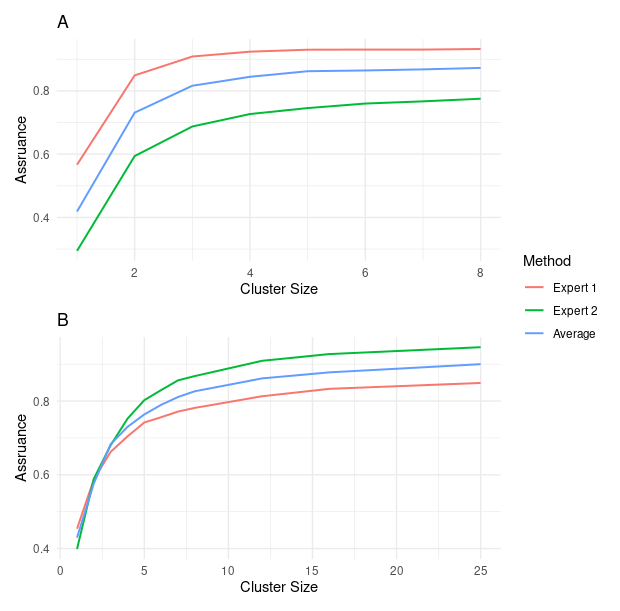}
\caption{(A) The assurance with different average cluster sizes for the time to thrombectomy. Expert 1 is more optimistic about the result than Expert 2 since the assurance for any chosen average cluster size is larger. (B) The assurance  with different average cluster sizes in the case of the thrombectomy rate. Expert 1 is more pessimistic about the result than Expert 2 in this case, and we can see Expert 2 and the average do reach an assurance of 0.9 in the plot, while for Expert 1 the assurance with average cluster sizes of 25 is around 0.87.}
 \label{fig:ass}
\end{figure}

The estimated sample size using the design prior distributions for expert 1, expert 2 and the average were 150 in each case, based on a minimum assurance of 90\%. This is due to the fact that the experts were very optimistic about the improvement in time to thrombectomy in the treatment arm, represented by $\delta$, with almost all of the prior mass in each case being above zero in Figure \ref{fig:den}. We note that this sample size is much smaller than that from the original power calculation, assuming an MCID of 30 minutes, of 564. To investigate the relationship between the assurance and the cluster (and hence sample) size, we instead use 50 clusters. The results are given in Figure \ref{fig:ass}(A). 

We see that in each case the assurance, like power, is an increasing function with cluster size. Expert 1 is most optimistic about the treatment, with expert 2 less optimistic and the average lying somewhere between the two. As the cluster size gets very large each assurance curve will tend to the probability, under that expert's design prior distribution, that the treatment effect is positive.

\subsubsection{Assurance for the thrombectomy rate}
\begin{figure}[H]
    \centering
\graphicspath{ {./images/} }
\includegraphics[width=\linewidth, height= 6cm]{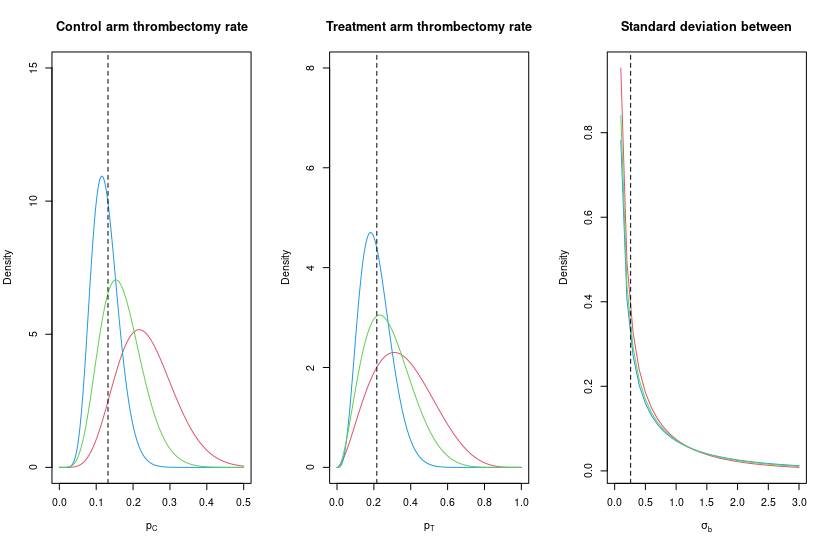}
\caption{The elicited prior probability density functions of the parameters ($\lambda, \delta, \sigma_b $) for the thrombectomy rate. Expert 1 is given in red, expert 2 in blue and the average in green. The vertical dashed lines are the values used in the original power calculation.}
 \label{fig:dbin}
\end{figure}

The elicited design prior distributions associated with the thrombectomy rate from expert 1, expert 2 and the average are given in Table \ref{tab:1} and plotted in Figure \ref{fig:dbin}. We use these to calculate the assurance, and hence sample size, with a target assurance value of 90\%. In this case we used INLA for our assurance and sample size calculations \textcolor{black}{based on the general conclusions from the simulation in Section \ref{Sec:comp}}.

The estimated sample sizes using the design prior distributions from expert 1, expert 2 and the average are 4800, 1650 and 3150 respectively. In this case the experts were relatively pessimistic about the likely values of the treatment effect for the thrombectomy rate, relative to the sample size estimate from the power calculation of 894. However, the required primary analysis population to ensure an adequate sample size for the time to thrombectomy outcome means that there will need to be between 2600 and 4300 patients recruited to the trial, meaning that in practice both expert 2 and the average will likely achieve 90\% assurance, and expert 1 will achieve relatively high assurance. Assurance also has a different interpretation to power, and so there is no reason why matching the values of power and assurance is an equivalent exercise. 

We have produced a plot of the assurance for different average cluster sizes, based on the 150 clusters in SPEEDY, for both experts and the average, and this is given in Figure \ref{fig:ass}(B).

We see a similar scenario as in the continuous outcome case, with the assurance increasing for increasing numbers of patients in each cluster. The main difference is in the ordering of the curves, with expert 2 providing the highest assurance for each cluster size and expert 1 providing the lowest, whereas in Figure \ref{fig:ass}(A) this was the opposite way round.

\section{Conclusions}\label{Sec:sum}

In this paper we have considered the problem of choosing the sample size, using a Bayesian approach, for a two-arm superiority cluster RCT with a continuous outcome and a binary outcome. We have compared the inference using MCMC to INLA based on appropriate mixed effect models. From the comparison we found that the use of INLA has advantages in cluster randomized trials with Bayesian designs, as it was as accurate as MCMC with a large number of MCMC samples ($K=10,000$ or more), but was typically faster to implement compared to MCMC, especially when the trials requires a large sample size in the continuous case, and in general in the binary outcomes case.

We used the SPEEDY trial as a case study of the sample size choice via an assurance calculation, as SPEEDY has 2 primary outcomes: both a continuous and a binary outcome. In the original sample size calculation SPEEDY did not consider the uncertainty in the model parameters, and so we performed an expert elicitation to specify suitable design prior distributions for the parameters. The expert elicitation was performed with two experts, and we calculated the assurance, and hence sample size, for each expert separately and the average of both experts. The findings were that the assurance and resulting sample sizes were smaller than with the original power calculation for the continuous case, since both experts were relatively optimistic about the ability of the SPEEDY pathway to reduce the time to thrombectomy by more than the values used in the power calculation, whereas the resulting sample sizes were much larger than their values from the power calculations for the binary outcome, as both experts felt that the value used for power in this case was fairly ambitious. Due to the nature of trial, with the two outcomes needing to be powered simultaneously, the actual sample size for the binary outcome realised in SPEEDY will provide high assurance for both experts. 

\textcolor{black}{In general, assurance is particularly beneficial when there is substantial uncertainty in the values of nuisance parameters to which the power calculation is sensitive. One such parameter considered in this paper is the ICC in a cluster randomised trial, which can be particularly challenging to estimate accurately a priori. Assurance provides a way to take this uncertainty into account, and provides a sample size which is more robust to mis-specification than a power calculation using a single estimated value. Trial statisticians should consider using assurance in place of power whenever they have substantial uncertainty about sensitive parameters in a power calculation.}

The calculation of the assurance and sample size for large trials, particularly with binary outcomes, would be almost prohibitively computationally expensive and time consuming given current widely available computing power without the use of INLA, as MCMC takes a very long time in these cases. This assurance approach, together with INLA (or MCMC) for inference, could be extended to more complex cluster randomized trial designs, including survival outcomes, longitudinal designs,  multi-arm trials and adaptive designs. This is left for future work.

\end{document}